\documentstyle[preprint,aps]{revtex}
\tighten
\draft
\widetext
\preprint{NYU-TH/00/03/09}
\bigskip
\bigskip
\begin{document}
\title{Cosmological Constant and Fermi-Bose Degeneracy}
\medskip

\author{Gia Dvali\footnote{E-mail: gd23@scires.nyu.edu}}

\bigskip
\address{Department of Physics, New York University, New York, NY 10003}

\bigskip
\medskip
\maketitle

\begin{abstract}
{} We study some cases in $D=4$ and in infinite-volume high dimensional
theories when unbroken supersymmetry in the vacuum cannot guarantee
Fermi-Bose degeneracy among excited states.  In 4D we consider an example
in which both supersymmetry and $R$ symmetry are unbroken in the vacuum, 
and the cosmological constant vanishes. However,
theory admits solitons that do not allow existence of conserved
supercharges. These objects are magnetically charged global monopoles
which in some respect behave as point-like particles, but
create a solid angle deficit which eliminates asymptotically covariantly
constant spinors and lifts Fermi-Bose degeneracy of the spectrum.
The idea is that in some "dual" description monopoles with
global topological number and gauge magnetic charge may be replaced
by electrically chaged particles with global Noether charges, e.g., such as
baryon number.
Alternatively theories with infinite volume extra dimensions may support
unbroken
bulk supersymmetry without Fermi-Bose degeneracy in the brane spectrum.
We suggest a scenario in which brane is a source similar to a
global monopole embedded in 3
infinite extra  dimensions. Brane produces a deficit angle at infinity and
may localize a meta-stable
4D graviton.  Although bulk cosmological constant is  zero, conserved
supercharges
can not be defined on such a background. 

\end{abstract}
\pacs{}

\subsection{Introduction.}

 Broken supersymmetry is hard to reconcile with vanishing of the cosmological
constant. In the presence of gravity even unbroken supersymmetry
alone does not suffice to guarantee the zero vacuum energy, but
under reasonable assumptions, e.g. such as $R$ symmetry, this can be done.
In this view perhaps the way for understanding the vanishing
of the vacuum energy, is to look for the theories in which unbroken
supersymmetry in vacuum cannot guarantee Fermi-Bose degeneracy
among excited states.
One may think of  two possible strategies in this direction. One is to
search for non-degenerate states directly in four-dimensional
theories, generalizing earlier observation by Witten in $2 + 1$
dimensions\cite{witten}.

 Alternatively \cite{dgp}\cite{witten1},  Fermi-Bose non-degeneracy may be
compatible  with unbroken bulk supersymmetry, if matter is localized on  a
brane embedded in {\it infinite-volume}
extra dimensions. In this way cosmological constant may be controlled by
bulk supersymmetry, while supersymmetry may be completely broken on the
brane.  Example of infinite-volume extra dimensions was recently suggested
in \cite{grs}. In this framework four dimensional Newtonian gravity is
reproduced by a meta-stable graviton localized
on the brane\cite{grs,quasi,dgp}. 

In this letter we consider both possibilities. First we shall study some
candidate solitonic states that make impossible the definition of unbroken
supercharges in four-dimensions,
although supersymmetry and $R$-symmetry
are unbroken in vacuum.  These objects are global
monopoles that are known to create a solid angle deficit at
infinity. Due to this solid angle deficit, the covariantly constant spinors
do not exist on such a background and there are no conserved supercharges.
Thus there is no reason for Fermi-Bose degeneracy in the monopole spectrum.
In many respect the dynamics of the global monopoles is reminiscent of
the one of the point-like particles with a finite mass localized in the
tiny core. We show that this particles can carry gauge quantum numbers and in
particular both Abelian or non-Abelian magnetic charges.
In particular, with an appropriate choice of a ``magnetic''
gauge group $H_{magnetic}$,
the magnetic charge assignment of the global monopoles can be made
identical to the electric Yang-Mills charges of the standard
model particles under $H_{electric} = SU(3)\otimes SU(2)\otimes U(1)$,
much in the
same way as for the gauge monopoles in $SU(5)$, as observed by Vachaspati
\cite{tanmay}. However, we shall
not restrict $H_{magnetic}$ by such a choice.
In this way we end up with the objects that in certain sense are ``dual''
to ordinary particles. Instead of global Noether charge
(baryon or lepton charge), they carry global topological charge
(winding number), instead of electric charges (color and ordinary electric
charge), they carry $H_{magnetic}$-magnetic charge, etc.
An interesting fact about these states is that despite the unbroken
supersymmetry in the vacuum, they are not Fermi-Bose degenerate.
So the theory describes a toy Universe with zero cosmological term,
but non-equal masses of fermions and bosons. The hope for more realistic
model-building is that in some dual description solitons may be replaced by
particles (topological charges with baryon or lepton numbers, etc..)
which are not Fermi-Bose degenerated, but cosmological constant is still
zero due to supersymmetry.

Secondly we consider infinite volume theories and suggest some new ways of
quasi-localization of gravity without invoking negative tension
branes. One possibility is that our brane originates from spontaneous
breaking of global $O(3)$ symmetry in theories with three extra
dimensions and has the structure of global monopole in transverse space.

Such an object produces a deficit angle at infinity and in a certain
limit may support a meta-stable graviton state. 
Since the brane appears as a solution of an
underlying sigma model on a background with zero bulk cosmological
constant, it may avoid problems with violation of the weak energy positivity
conditions pointed in\cite{witten1} (see also \cite{quasi}).

\subsection{Supersymmetry in the Global Monopole Background
in Four Dimensions.}

 In the inspiring paper \cite{witten} Witten made an observation
that unbroken supersymmetry (and thus zero cosmological constant)
is not incompatible with Fermi Bose
non-degeneracy in $2 + 1$ dimensions.

The idea  is roughly as follows. In supergravity theories the supercurrent
in general is not conserved in the usual sense,  but rather is covariantly
conserved
\begin{equation}
D_{\mu}J^{\mu} = 0
\end{equation}
However,  in the presence of  a covariantly constant spinor
\begin{equation}
D_{\mu}\epsilon = 0
\label{spinor}
\end{equation}
the conserved current can be constructed
\begin{equation}
\bar{\epsilon}J^{\mu}
\end{equation}
and thus one can define a globally conserved supercharge
\begin{equation}
Q = \int dx^3 \bar{\epsilon}J^0
\end{equation}
Now in three dimensions any localized mass is known to produce a conical
geometry at infinity\cite{des}. Such a geometry makes in general impossible
the existence of covariantly constant spinors and thus of unbroken
supercharges\cite{hen}.\footnote{In the presence of gauge fields, 
Killing spinors may still exist, since the deficit angle can be
compensated by an Aharonov-Bohm phase.\cite{vortex}}. Thus although 
supersymmetry is
unbroken in the vacuum and vacuum energy vanishes, there is no Fermi-Bose
degeneracy among the excited states.  Explicit examples along Witten's idea
were
considered in some papers
\cite{vortex},\cite{examples}.
Needless to say it would be very important to find some
sort of generalization of this effect to four-dimensions. 

In this letter we will consider some possible candidates that may
generalize the three-dimensional behavior of point masses to four dimensions.
We shall look for $N=1$ $D=4$ supergravity theories, in which both
supersymmetry and $R$ symmetry are unbroken in the vacuum, but yet there
is no Fermi-Bose degeneracy among certain excited solitonic states.

 The model consists of four chiral superfields. Three of them,
$\Phi_a,~~a=1,2,3$ compose a triplet under an internal symmetry
group $O(3)$, while the fourth superfield $X$ is a singlet.
The superpotential is given by
\begin{equation}
W = X(\Phi_a^2 - v^2)
 \end{equation}
where $v$ is a real mass parameter and for simplicity the coupling
constant
is set to one. This theory has a global $O(3)$ invariance, plus
an $U(1)_R$ R-symmetry, under which $X$ transforms in the same way as $W$,
whereas $\Phi_a$-s are invariant.

 The vacuum of the theory is given by (below we shall denote chiral superfields
and their scalar components by the same symbols)
\begin{equation}
 X = 0, ~~~\Phi_a^2 = v^2
\end{equation}
Thus $O(3)$ is broken spontaneously to $O(2)$, whereas both
supersymmetry and $R$ symmetry are unbroken. This ensures that vacuum
energy
is zero to all orders in perturbation theory.

 Due to a nontrivial topological structure, however, this theory admits
topological knots. These knots are global monopoles, which in spherical
coordinates can be given by the solution
\begin{equation}
X=0, ~~~\Phi_a = f(r){r_a \over r}
\label{monopole}
\end{equation}
where $r_a$ are the components of the radius-vector ${\bf r}$
and $f(r)$ is a smooth
function such that
\begin{equation}
f(0) = 0, ~~~ f(r)|_{r \rightarrow \infty} \rightarrow v
\end{equation}
As shown by Barriola and Vilenkin \cite{vilenkin}, with this ansatz, 
the metric takes the following form
\begin{equation}
ds^2 = - a^2 dt^2 + b^{2}dr^2 + r^2d\Omega^2
\end{equation}
where asymptotically
\begin{equation}
 a^2 = b^{-2} = \left (1 - {v^2 \over M_P^2} - {2M_{core} \over M_P^2r}\right )
\end{equation}
where $M_P$ is the reduced Planck mass and $M_{core} \sim v$ 
is an integration constant, which is a negative
quantity\cite{monop}. Due to this fact the Newtonian potential of the
monopole core is repulsive. This metric describes a
space with
a solid angle deficit $4\pi {v^2 \over M_P^2}$. The
$\theta = \pi/2$ surface is a cone with a deficit
angle $\delta = 2\pi v^2/M_p^2$.

 Of our interests are the supersymmetric transformations in
the monopole background. Choosing the vierbein as
\begin{equation}
 e_a^{\mu} = diag ({1 \over a}, a, {1\over r}, {1 \over r {\rm sin}\theta}),
~~a^2 = \left (1 - {v^2 \over M_P^2}\right )
\end{equation}
(where $\mu = t, r, \theta, \phi$)
the only non-vanishing components of the spin connection are
$\Gamma_{\theta} = - {a\over 2}\gamma_1\gamma_2$ and
$\Gamma_{\phi} = - {a \over 2}{\rm sin}\theta \gamma_1\gamma_3
- {1 \over 2}{\rm cos}\theta \gamma_2\gamma_3$. On such a background,
the conditions for Killing spinors
\begin{eqnarray}
&& \delta\psi_{\theta} = (\partial_{\theta} +
{a\over 2}\gamma_1\gamma_2)\epsilon = 0 \\
&& \delta\psi_{\phi} = (\partial_{\phi} + 
 {a \over 2}{\rm sin}\theta \gamma_1\gamma_3 +
{1 \over 2}{\rm cos}\theta \gamma_2\gamma_3)\epsilon = 0
\end{eqnarray}
can not be satisfied and no conserved supercharges can be 
defined to control Fermi-Bose degeneracy
in the monopole spectrum.

Alternatively, this can be seen in the language of zero modes.
In the global supersymmetry  limit the zero mode in the
monopole
background is
\begin{eqnarray}
 && \delta\psi_{X+} = \sqrt{2}(f^2(r) -v^2)\epsilon_+\\
 &&   \delta\psi_{+}^a = -i \sqrt{2}
\partial_{\mu}\Phi^a \gamma^{\mu}\epsilon_-,
\end{eqnarray}
where $\epsilon_{\pm}$ is an eigenspinor of $\gamma_5$. Note that
$\delta\psi_{X+}$ and ${r_a\over |r|}\delta\psi_{a+}$
are normalizable,
whereas $\delta\psi_{\theta}$ and $\delta\psi_{\phi}$ are not.
The reason behind this is the following. We can say that global
supersymmetry  is spontaneously broken
in the monopole background. The order parameters for this breaking are the
auxiliary ($F$) component of the chiral superfield $X$
\begin{equation}
\langle F_X \rangle = (f(r)^2 - v^2)
\end{equation}
and also an auxiliary ($D$) component of a composite real superfield
$\Phi_a^*\Phi_a$
\begin{equation}
\langle (\Phi_a^*\Phi_a )_D \rangle = |\partial_j\Phi_a|^2
\end{equation}
Since both $F_X$ as well as radial derivatives vanish  away from the core
at least as $1/r^2$, the corresponding fermion variations
$\delta \psi_X$ and
${r_a\over |r|}\delta\psi_a$ are normalizable. Away from the core breaking is
dominated by angular gradients, which only drop-off as inverse square of
distance and goldstino is not normalizable. In fact this
is also to be expected from the fact that monopole configuration
spontaneously breaks both internal symmetry as well as
coordinate rotations and leaves unbroken  an $O(3)$ symmetry
of combined space and internal
rotations, which leaves invariant a product $r_a\Phi_a$. 
Non-normalizable fermionic zero modes are simply the supersymmetric
partners of the Nambu-Goldstone fields corresponding to this breaking.
Monopole configuration interpolates between the core where internal
$O(3)$ is restored, but coordinate symmetry and supersymmetry are broken
maximally,  to the outer region where the strength of supersymmetry
and coordinate symmetry breakings weakens but internal symmetry is
maximally broken.
Since both coordinate and supersymmetries are broken by
angular gradients which vanish
away from the core, all energy densities must vanish as $\sim
1/r^2$.  Although locally energy density
can be arbitrarily small, in the background of the global monopole the
globally conserved supercharge
linearly diverges together  with the total energy of the configuration 
\begin{equation}
 Q(\epsilon) \sim 4\pi \bar{\epsilon}\gamma_0\epsilon v^2 R
\end{equation}
where $\epsilon$ is an arbitrary constant spinor and $R$ is the
distance from the
core of the monopole. 
In the case of supergravity, the zero mode acquires an additional
non-normalizable gravitino component
$\delta \psi_{\mu}$ which renders it unphysical.

\subsection{Monopoles as Particles.}

The global monopoles due to the divergent energy,
are different from the ordinary localized sources.
However, in many respect they can behave like objects with a
finite mass localized in the core.
We can treat the cores of the global monopoles
as point-like objects of finite mass moving in the ``clouds'' of the Goldstone
gradient
energy produced by their Goldstone field. In this way divergence in the
total energy of the configuration is not important for the core dynamics,
in a same way as the total infinite energy of the matter in the Universe 
is not important for the particle interactions at short distances.

 Of course, the precise numbers do not work as well as in the case of the 
ordinary
matter. For instance, roughly $n \sim 10^{19}$ global monopoles,
with the core mass equal
to the proton mass, presented within the observable part of the Universe
would be enough to produce an energy density equal to the critical
\footnote{This estimate assumes that all the monopoles within the observable
part have the same charge (like baryons). If one assumes the equal number
of monopoles and anti-monopoles within the Hubble horizon,
the allowed number would be much higher.}.
However, we shall ignore this complications, since our goal is to 
understand whether in principle $4D$ theories with unbroken vacuum 
supersymmetry and Fermi-Bose non-degeneracy are possible.
We see no reason to think that the ``dual'' Universe in which particles
are described by monopoles must obey similar cosmological constraints.

 To understand how far the analogy between particles and
the global monopoles can
be extended, we can estimate their
interaction potential due to various forces.

We shall discuss gravitational effects first. The leading contribution
comes from a tree-level one-graviton exchange
\begin{equation}
\sim G\int dx^4~ d^4x'~
T_{\mu\nu}(x)~ G^{\mu\nu\alpha\beta}(x - x')~T'_{\alpha\beta}(
x') ~,
\label{exch}
\end{equation}
where $G$ is a Newtons constant and the graviton propagator is given by
\begin{equation}
G^{\mu\nu\alpha\beta}(x - x') = \int {dp^4\over (2\pi)^4} {{1\over
2}(\eta^{\mu\alpha}\eta^{\nu\beta} + 
\eta^{\mu\beta}\eta^{\mu\alpha}) - {1\over
2}\eta^{\mu\nu}\eta^{\alpha\beta}  \over p^2 - i\epsilon} e^{-ip(x - x')}
\end{equation}
For the rough estimate we shall ignore the effect of the curvature
(which dies-off as $\sim 1/r^2$ away from the core)
and use the flat space propagator.
In the first approximation, we can assume that the structure of
gravitating sources are not affected by simultaneous presence of two monopoles,
and ignoring the effect of the core we can set:
\begin{equation}
 T_t^t = T_r^r = {v^2 \over |{\bf r}|},~~
 T_t^{'t} = T_r^{'r} = {v^2 \over |{\bf r'} - {\bf R}|}
\end{equation}
and all other components zero. Here $R$ is a distance
between the monopoles.
In this case (\ref{exch}) is zero. So the gravitational
interaction of monopoles will be governed by the Newtonian interactions
between the two cores\footnote{We shell ignore the tidal acceleration
$\sim 1/r^2$, which may be 
very important for non-static sources moving with a nonzero
impact parameter\cite{tidal}.}
\begin{equation}
 V(r) \sim G{M^2_{core} \over r}
\label{coregravity}
\end{equation}
This is sub-dominant with respect to the force mediated by the Goldstone
field. For a monopole-anti-monopole pair the Goldstone-mediated
interaction gives an attractive constant force
\begin{equation}
V(r) \sim v^2r
\end{equation}
However, we are more interested in monopoles of the same charge, since
a monopole-anti-monopole pair produces no deficit angle at infinity.
Below we shall assume that vacuum topology is such that monopoles with
the higher winding numbers
are possible.
In this case the long range interaction is repulsive (there still however
can be an attractive short-range interaction between the cores).
Due to this the $n$-monopole system is unstable. This fact can be understood
qualitatively as follows. Consider $n$ monopoles uniformly 
distributed inside the sphere
of radius $R$. For distances $r>>R$ the configuration of the Higgs field is
similar to the one of the monopole with winding number $n$, and gradient
energy diverges as
\begin{equation}
 E_{out} \sim n^2v^2(r_{max} - R)
\end{equation}
while inside the sphere it scales as
\begin{equation}
 E_{in} \sim nv^2R
\end{equation}
Therefore it is energetically favorable for $R$ to grow.

 However, the linear potential between monopoles can be modified
in many cases.
In particular, it is easy to imagine situation when this force
vanishes because of the complex
structure of the monopole. For instance,
imagine that there are two broken global symmetries
$G$ and $G'$ that produce monopoles. Let $\Phi$ and $\Phi'$ be the Higgs fields
responsible for such breakings. We assume for simplicity that
$\Phi$ is trivial under $G'$ and so is $\Phi'$ under $G$ respectively.
Then the two fields produce independent monopoles which we shall refer to
as
$\Phi$-monopoles and $\Phi'$-monopoles respectively.
If there is a $G\otimes G'$-invariant contact interaction in the
potential
\begin{equation}
 - \Phi^*\Phi\Phi^{'*}\Phi'
\end{equation}
the monopoles of the two sorts will tend to create the bound-states.
Such interaction can be easily arranged by introducing the following couplings
in the superpotential
\begin{equation}
W =  Y(h\Phi^2  - h'\Phi^{'2}) + ...
\label{higgsinteraction}
\end{equation}
Where $Y$ is a chiral superfield and $h,h'$ are constants
The equation of motion for $Y$ then forces $h\Phi^2 =  h'\Phi^{'2}$
and encourages $\Phi$ and $\Phi'$
to have coincident zeros. Thus $\Phi$-monopoles get bounded
to $\Phi'$-monopoles. The strength of the binding force
is set by the parameters $h,h'$ and the magnitude of the two VEVs.
Note that the binding interaction
does not distinguish between monopoles or anti-monopoles in the opposite
sectors. So formation of a monopole-monopole ($MM$) boundstate
is as probable as
the formation of a monopole-anti-monopole ($M\bar M$) one. 
Depending on the parameters of the theory, the Goldstone-mediated
force between $MM$ and $M\bar M$ states may be either repulsive,
attractive or zero. In the latter case the interaction will be dominated
by a gravitational potential from the core.

 In addition global monopoles can have long-range gauge interactions.
In particular, as we shall show below, this long range interaction
can be due to their magnetic charge.

\subsection{Global Monopoles with Magnetic Charges.}

 Interestingly the global monopoles can carry magnetic charges
under both Abelian or non-Abelian gauge groups. 
Let us consider first the Abelian case. For this, in addition to the
spontaneously broken $O(3)$ global symmetry we introduce an
unbroken $U(1)_m$ gauge symmetry in the theory. The question is whether
a global monopole given by (\ref{monopole}) can acquire a $U(1)_m$-magnetic
charge. To answer this let us define an invariant two form
\begin{equation}
 M_{\mu\nu} = \epsilon_{abc}\Phi^a\partial_{\mu}\Phi^b\partial_{\nu}\Phi^c
\label{twoform}
\end{equation}
which determines a topological winding number
\begin{equation}
n = {1 \over 4\pi} \int_{S_2}  {M_{\mu\nu} \over |\Phi|^3} dx^{\mu}dx^{\nu}
\end{equation}
where integral is taken over a two-sphere surrounding the monopole.
Let us now couple this two-form to a gauge-invariant $U(1)_m$-field strength
\begin{equation}
 {\lambda \over 2} M_{\mu\nu}F^{\mu\nu}
\label{formcoupling}
\end{equation}
We shell treat this term as the perturbation on the stable monopole
background. 
The equation of motion then gives
\begin{equation}
\partial_{\mu}F^{\mu\nu} = \lambda \partial_{\mu}M^{\mu\nu}
\end{equation}
This tells us that $U(1)_m$-magnetic field $({\bf B})$ is radial, but does not
fix its magnitude. The magnitude is determined by minimizing the
$U(1)_m$-magnetic energy of the system
\begin{equation}
 E_{magnetic} = - \lambda {\bf M}{\bf B} + {{\bf B}^{2} \over 2}
\end{equation}
where ${\bf M_i} = \epsilon_{ijk} M_{jk}$. The energy is
minimized by
\begin{equation}
{\bf B} = \lambda v^3 {{\bf r} \over r^3}
\end{equation}
In this way, the global monopole acquires $U(1)_m$ magnetic charge
\begin{equation}
Q_m = {1 \over 4\pi} \int_{S_2}  F_{\mu\nu} dx^{\mu}dx^{\nu} = Nv^3\lambda
\end{equation}
proportional to the topological charge of its own. For this configuration
not to be singular $U(1)_m$ must be embedded in some non-Abelian group.
To do this, let us, instead of $U(1)_m$, introduce
$SU(N)$ gauge symmetry, spontaneously broken to
$SU(N-M)\otimes SU(M)\otimes U(1)$ by a Higgs field $\Sigma$ in the adjoint
representation. We can define a gauge-invariant two-form
\begin{equation}
 F_{\mu\nu} = {\rm Tr}\Sigma {\bf F}_{\mu\nu}
\label{nonabelian}
\end{equation}
where ${\bf F}_{\mu\nu}$ is the $SU(N)$ field-strength.  Then the coupling 
of the form (\ref{formcoupling}) between (\ref{nonabelian}) and
(\ref{twoform}) will induce a magnetic charge of the global monopole.
Effectively what happens is that $SU(N)$-gauge t'Hooft-Polyakov monopole
gets trapped in the core of the global monopole.

 Finally in case when gauge sector permits topologically stable monopoles
(just like in above $SU(N)$ example), the global monopole can acquire
magnetic charge by simply trapping the gauge monopoles inside its core 
due to a short-range Higgs interaction of the form similar to 
(\ref{higgsinteraction})
\begin{equation}
W_{cross} =  Y(h\Phi^a\Phi^a  -  h'Tr \Sigma^2)
\label{higgsinteraction1}
\end{equation}
This interaction tries to bring zeros of the two
Higgs fields together and thus
creates a boundstate of global and gauge monopoles.
For instance, we can choose $SU(2)$ gauge symmetry broken by triplet
$\Sigma^a$ and set its self-interaction term in the superpotential
\begin{equation}
W_{\Sigma} =  Z(Tr \Sigma^2  -  v_{\Sigma}^2)
\label{higgsinteraction1}
\end{equation}
where $Z$ is an additional superfield.
The asymptotic form of the solutions for the ``composite'' monopole
is
\begin{equation}
Y = Z = 0,~~
\Phi^a = \pm  \sqrt{{h' \over h }}\Sigma^a = v_{\Sigma}{r^a \over r},~~ 
A^{a}_{\mu} = \epsilon_{\mu a b}{r^b \over r^2}
\end{equation}
and has the same magnetic charge as the elementary $SU(2)$ monopole.

\subsection{Some speculations.}

 As we have seen, the global monopoles in $N=1$ $D=4$ supersymmetric theories
exhibit some interesting properties. Due to the solid angle deficit at
infinity they do not permit existence of conserved supercharges
and, thus, there is no {\it a priory} reason for a Fermi-Bose
degeneracy in the spectrum, although supersymmetry is unbroken and
cosmological constant vanishes.
 Despite the fact that the total energy of the configuration
diverges, in some respect they behave like point-like particles of
a finite mass localized at the core. These localized masses can carry
gauge quantum numbers and, in particular, both Abelian as well as non-Abelian
magnetic charges. If this toy picture can have any 
relation with the observed smallness of the cosmological term,
we should expect that
in some ``dual'' theory these monopoles are related
to the ordinary particles carrying electric Yang-Mills charges. 

It has been shown \cite{tanmay} that the $SU(3)\otimes SU(2) \otimes U(1)$
magnetic charges of the gauge monopoles produced in the symmetry breaking
\begin{equation}
SU(5) \rightarrow SU(3)\otimes SU(2) \otimes U(1)/Z_6
\label{gauge}
\end{equation}
are in correspondence with the electric gauge charges of the standard
model fermions. This interesting observation naturally
leads to the idea of the dual standard model\cite{tanmay}.

 We have shown that global monopoles can carry
any magnetic charge that can be carried by the local ones.
Then the following toy scenario emerges. We take $N=1$ supersymmetric theory
with a symmetry group $G_g\otimes G_l$, where $G_g$ is global and
$G_l$ is gauged. We assume that these symmetries are spontaneously broken
to $H_g\otimes H_l$ by Higgs fields $\Phi$ and $\Sigma$ respectively.
The breaking is such that $\pi_2(G_g/H_g)\neq 0$. This ensures 
the existence of stable global $\Phi$-monopoles.
On the other hand $\pi_2(G_l/H_l)$ may or may not be empty, since as we
have seen, the global monopoles can acquire magnetic charges even if the
vacuum of the broken gauge symmetry is topologically trivial.
If however, also  $\pi_2(G_l/H_l) \neq 0$, then there will
be local monopoles formed by $\Sigma$. If $G_l$ is simply connected,
then gauge monopoles are classified by non-contractable paths in $H_l$
and carry different magnetic charges. There can be a finer classification
\cite{fine} according  to the representations of a dual magnetic group.

 Now the global monopoles can acquire magnetic charges under $H_l$
via one of the above
discussed mechanisms, either by ``kinetic'' mixing (\ref{nonabelian})
or by forming the
boundstate due to the contact Higgs interaction 
(\ref{higgsinteraction}). In the latter case for each stable gauge monopole
there will be a composite stable global monopole with identical gauge charge.
These objects carry a global topological charge (``dual baryon number'')
and a local magnetic charge (``dual electric charge'').
\footnote{If $H_l = SU(3)\otimes SU(2) \otimes U(1)/Z_6$ then according to
\cite{tanmay} the magnetic charges of these monopoles will match the
$SU(3)\otimes SU(2) \otimes U(1)$-electric charges of the 
standard model fermions. This choice however is not necessary since the
dual ``electric'' group $\tilde {H_l}$ need not be isomorphic to $H_l$.}
They move in a cloud of Goldstone field which creates a solid angle 
deficit at infinity. No conserved supercharges can be defined on such a
background. On the other hand supersymmetry and $R$ symmetry
are unbroken in the vacuum and cosmological constant vanishes.
Then one may expect that in the dual picture this monopoles are replaced by
electrically charged particles (``baryons'' and ``leptons'') of
some more conventional theory
in which fermions and bosons will not be degenerate but vacuum energy will
be zero because of (hidden) unbroken supersymmetry.
Such a duality probably has to exchange the global topological
winding number with the global Noether charge (baryon or lepton numbers),

\subsection{Infinite-Volume Extra Dimensions.}

An alternative way of controlling the cosmological constant by
supersymmetry may be by going to a ``brane world'' scenario. In the standard
brane world picture the ordinary matter is localized on a brane embedded
in $N$ large extra dimensions. These can be as large
as millimeter with the fundamental Planck mass $(M_{Pf})$ around
TeV\cite{add}. The volume of extra dimensions is finite, either because
these dimensions are compact or because
the warp factor decays exponentially fast away from the
brane\cite{rs}. The compactification combined with non-trivial warp factors
in co-dimension two spaces were  considered\cite{cp}.
Compactification may take place
due to singularity at
finite distance from the brane as, for instance, in
\cite{ck}. In either case
there is an upper bound around $\sim$ mm on the volume of extra dimensions
from gravitational measurements. We shall define this volume as
\begin{equation}
 V_{extra} = \int dx^N_{extra} \sqrt{-g_{extra}}g^{00}(x_{extra})
\end{equation}
where the integration is taken over
an $x_{extra}$-dependent part of the metric.
This volume factor sets the normalization of the 4D zero mode graviton and
thus the relation between fundamental and observable Planck scales
\begin{equation}
M_P^2 = M_{Pf}^{2+N}V_{extra}
\end{equation}
In this set-up, it is very difficult to control the value of the cosmological
constant by bulk supersymmetry.  Indeed, since the extra volume is finite,
the extra coordinates can be integrated over and at large distances, the
effective
four-dimensional description should be valid. Then by four-dimensional
general covariance the brane and bulk states should be gravitationally
connected through the zero mode graviton.
Now, supersymmetry must be broken among the brane states at least at scale
$TeV$.
This breaking then will universally get transmitted to all bulk modes by
gravity.
This transmission is at best volume suppressed, so that by
dimensional
argument the mass splitting among the bulk states is at least  $\Delta
m^2\sim TeV^4/M_P^2$
and the resulting cosmological constant is at least $\sim \Delta m^2
\Lambda^2_{cut-off}$.

The crucial point is that in theories with infinite $V_{extra}$ there is
no {\it a priory} reason for the above inconsistency \cite{dgp,witten1}: since
the volume is infinite, the theory is never four-dimensional and effective 4D
description is never valid. In other words, supersymmetry broken on the
brane may not get transmitted in the bulk. So the high-dimensional theory
can be supersymmetric even though brane observers will not see any
Fermi-Bose degeneracy.

An interesting model of infinite-volume extra dimensions was invented in
\cite{grs}
as modification of RS scenario\cite{rs}. (Modification of gravity
at large distances was suggested earlier in \cite{ian}.)
In their case the warp factor in
five dimensional metric
\begin{equation}
ds^2 = A(y) ds^2_4 - dy^2
\end{equation}
instead of vanishing as $y \rightarrow \pm\infty$, was asymptotically
approaching a tiny constant value. As a result the space is infinite and
asymptotically flat in $y$ direction.
This structure was achieved by expense of introducing a combination of
positive and negative tension branes. Despite this fact a correct
four-dimensional Newtons law
is reproduced at intermediate scales on the brane. As was shown in
\cite{grs,quasi,dgp},
this can be explained by the existence of meta-stable resonant graviton
localized on the brane.  There are two potential problems with this
scenario.  One, to be put aside in the present paper, is the fact that 4D
gravity on the brane is mimicked by massive spin-2 fields,
which can not reproduce the predictions of Einsteins theory \cite{dgp}.
\footnote{This is true whenever dominant contribution to 4D gravity comes from
a massive spin-2 state(s).
Some way out of this problem in the context of \cite{ian}
was suggested in\cite{ian1}}
\footnote{In \cite{dgp,dgp1} it was noted that the unwanted contributions
to one-graviton exchanges
may be canceled by unconventional states like ghost 
(we understand that this was essentially
confirmed by ref\cite{t55,rat}), but then
it is hard to make sense of the theory, even if this states only
dominate exchanges at finite distances. This goes in contrast to
the conclusions of \cite{quasi1} and \cite{grs1} in which it was
argued that such states can solve the problem (conclusion
about their long-distance behavior is opposite in these two references).
We think that existence of the physical ghost is a problem at any distances.
For instance, even if canceled in all one-particle exchanges,
it is not clear what
can prevent their appearance in the final state.}

Second is the violation of the weak energy condition, which has to do with
the existence of
AdS portion bounded by negative tension branes embedded in Minkowski
space\cite{witten1}.  In this respect it is very
important to obtain a source that quasi-localizes gravity as a solution of
the underlying theory.
We want to note that the branes which appear from spontaneous breaking of
non-Abelian global symmetry may have this property. We consider
co-dimension $3$ case. In this case the brane is a global monopole embedded in
three extra dimensions\footnote{
From a different perspective the higher co-dimension global 
defects were studied
by Vilenkin and Olasagasti\cite{alex1} and by Cho. I am grateful to these authors for
correspondence and discussions.}
Consider a theory  in $7$ dimensions with  a spontaneously broken global
$O(3)$ symmetry,
by the VEV of a triplet scalar field $\Phi_a$. The potential is
\begin{equation}
 V =  (\Phi_a\Phi_a - v^2)^2
\end{equation}
Due to topological arguments this theory has co-dimension $3$ objects,
independent of  four space-time coordinates, described by
eq(\ref{monopole}), where $r$ has to be understood as the radial coordinate
perpendicular to the brane.  The general spherically symmetric
ansatz for the metric can be written
\begin{equation}
ds^2 = a^2(r) \eta_{\mu\nu}dx^{\mu}dx^{\nu} - b^2(r)dr^2 -  r^2(d\theta^2 +
sin^2\theta d\phi^2)
\end{equation}
Much in the same way as for the global monopole, away from the core we can
set $f(r) = v^2$ and the only contribution to the source will come from
the angular derivatives.
As a result
\begin{equation}
 T_{\mu}^{\mu} = T_r^r = {v^2 \over r^2} + ...
\label{Tmunu}
\end{equation}
and all other components zero.
The ellipses stand for the sub-leading correction.
With this source,
there is a straightforward generalization of the Barriola-Vilenkin solution,
which (in $M_{fp}$ units) reads:
\begin{equation}
a^2 = b^{-2} = 1 - v^2 =\alpha
\end{equation}
As in the four-dimensional case, there is a deficit angle $4\pi v^2$.
In principle, by tuning the parameter $v$, one can make angular
deficit arbitrarily close to $4\pi$.  Now let us study the issue of
graviton localization. 

First let us ignore the subleading corrections in (\ref{Tmunu}). That means
assume that $a$ goes to it asymptotic constant value fast enough.
Then the volume of the transverse space
\begin{equation}
V_{extra} = 4\pi \int dr a^2br^2
\end{equation}
is infinite unless $a$ vanishes asymptotically. This volume has two
contributions
\begin{equation}
V_{extra} = V_{core} + 4\pi\int_{r_{core}}^{\infty} ar^2dr
\end{equation}
The second term diverges, unless we tune $\alpha = 0$ in which case the
volume is finite and is given by $V_{core}$. There is no reason to expect
in $\alpha \rightarrow 0$ limit  any pathological behavior in the
core. Assuming this the theory must have a finite volume and must support
a localized zero mode graviton in the core of the monopole.
The following fluctuation about the background metric
\begin{equation}
ds^2 = a^2(r) (\eta_{\mu\nu}  + h_{\mu\nu}) dx^{\mu}dx^{\nu} - 
b^2(r)dr^2 -  r^2(d\theta^2 +
sin^2\theta d\phi^2)
\end{equation}
is a zero mode. This mode is normalizable if $V_{extra}$ is finite.
Now, we can make $\alpha$ to be nonzero, but arbitrarily small. 
Now the volume becomes
infinite and the zero mode becomes non-normalizable. However, if $\alpha$ is
small enough, the system by continuity  should still allow a localized
meta-stable mode, with  an arbitrarily long life-time.

 A problem with this argument is that it relies on an oversimplified
anzats for the monopole core. For a realistic monopole, there are sub-leading
corrections $\sim {1 \over r^2}$ to eq(\ref{Tmunu}) (recall that
$(f^2 - v^2)$ dies away as ${1 \over r^2}$). So the volume for $r
\rightarrow \infty$ is divergent,
which makes existence of
quasi-localized mode unclear.
More precise conclusion depends on the model of monopole core
which will not be discussed here.

 Summarizing, in order to avoid the violation
of the weak energy condition one may try to quasi-localize gravity on
a brane which is a solution of an underlying sigma model.
Under certain assumption regarding the core structure, this brane may
supports a meta-stable graviton
because of the deficit angle at infinity.
The next step would be to supersymmetrize the initial theory in
such a way that
supersymmetry gets restored in the bulk. However, this is not so easy, since
due to the same deficit angle we will
not be able to define globally conserved 
supercharges in such a background. However, this is not necessarily
a problem, since as in four-dimensions, the bulk vacuum energy density
is controlled by the order parameters that break supersymmetry.
These are angular
gradient densities that die away from the core as $1/r^2$. Thus all the energy
densities in the bulk must be controlled by the distance from the brane
and get zero at infinity. Although the total transverse energy
is linearly divergent, this is not of any problem, since the four-dimensional
metric on the brane is flat.

 An alternative approach one can try is to reintroduce Killing spinors,
by adding
gauge fields and canceling the
deficit angle by Aharonov-Bohm-type phases, just as in the
three dimensional case of ref\cite{vortex}. In this way some of the
supercharges may be unbroken, but act trivially on the brane.

\subsection{conclusions.}

 It is not clear how the $4D$ scenario discussed in this paper can be related
to an observable Universe. Yet it has some ingredients for this connection
and is an example of $4D$ supersymmetric theory with vanishing cosmological
constant, in the background that makes impossible definition of unbroken
supercharges. Probably the way this framework may
be related to more conventional ones is through some sort of a duality,
which relates solitons with particles, magnetic charges with electric charges
and in addition global topological charges (winding number)
to conserved global Noether charges (such as baryon or lepton number).

 Among many open questions, not addressed in this letter,
there are cosmological issues.
For instance with a rough estimate the number of the global
monopoles that would contribute $\rho_{baryon}$ energy density
within the observable part of the Universe is much less than the
number of baryons. So one may ask how this fast fits in the above picture.

We do not see any reason why two theories related through duality should
have similar cosmological history. We expect that duality relates
spectra and quantum numbers, but not the actual number of occupied
states,  which is determined by the cosmological initial conditions.

 Alternatively infinite volume extra dimensional theories may provide
example of unbroken supersymmetry compatible with four-dimensional
Fermi-Bose non-degeneracy. We have suggested that gravitating
sources similar to global topological defect, producing
angular deficit in the extra space,
may under certain conditions localize $4D$ meta-stable graviton.

\vspace{0.3cm}

{\bf Acknowledgments}

\vspace{0.2cm} 

It is pleasure to thank B. Bajc, G. Gabadadze,  E. Poppitz, M. Porrati,
M. Shifman,
T. Vachaspati and A. Vilenkin
for very useful discussions.
This work is supported in part by a David and Lucile  
Packard Foundation Fellowship for Science and Engineering.


\begin{references}

\bibitem{witten} E. Witten, Int. Jour. Mod. Phys. {\bf A10}
(1995) 1247,  hep-th/9409111.

\bibitem{dgp} G. Dvali, G. Gabadadze and M. Porrati, hep-th/0002190.

\bibitem{witten1} E. Witten, hep-ph/0002297.

\bibitem{grs} R. Gregory, V.A. Rubakov and S.M. Sibiryakov, hep-th/0002072.

\bibitem{quasi} C. Csaki, J. Erlich and T.J. Hollowood, hep-th/0002161.

\bibitem{tanmay} T. Vachaspati, Phys. Rev. Lett. {\bf 76} (1996) 188.

\bibitem{vortex} K.Becker, M. Becker and A. Strominger, Phys. Rev. {\bf D51}
(1995) 6603; 

\bibitem{des} S. Deser, R. Jackiw, and G.'t Hooft, Ann. Phys. (N.Y.)
{\bf 152} (1984) 220.

\bibitem{hen} M. Henneaux, Phys. Rev. {\bf D29} (1984) 2766.

\bibitem{examples} S. Forste and A. Kehagias, hep-th/9610060.

\bibitem{vilenkin} M. Barriola and A. Vilenkin, Phys. Rev. Lett. {\bf 63}
 (1989) 341.

\bibitem{monop} D. Harari and C. Lousto, Phys. Rev. {\bf D42} (1990) 2626.


\bibitem{tidal} W.A. Hiscock, Phys. Rev. Lett. {\bf B64} (1990) 344.

\bibitem{add} N. Arkani-Hamed, S. Dimopoulos and G. Dvali, 
Phys. Lett. {\bf B429} (1998) 263; Phys. Rev. {\bf D59} (1999) 086004.


\bibitem{fine} P. Goddard, J. Nuyts, and D.I. Olive, Nucl, Phys.
{\bf B125} (1977) 1.
 
\bibitem{rs} L. Randall and R. Sundrum, Phys. Rev. Lett. {\bf 83} (1999)
3370.

\bibitem{cp} A. Chodos and E. Poppitz, Phys. Lett. {\bf B471} (1999) 119,
hep-th/9909199.

\bibitem{ck} A. G. Cohen and D. B. Kaplan, Phys. Lett. {\bf B470} (1999)
52, hep-th/9910132.

\bibitem{dgp1} G. Dvali, G. Gabadadze, M. Porrati, hep-th/0003054. 

\bibitem{ian}  I.I. Kogan, S. Mouslopoulos, A. Papazoglou, G. G. Ross,
J. Santiago, hep-ph/9912552 

\bibitem{ian1} I.I. Kogan, G.G. Ross, hep-th/0003074.

\bibitem{t55} G. Kang, Y.S. Myung, hep-th/0003162.

\bibitem{rat} L. Pilo, R. Rattazzi, A. Zaffaroni, hep-th/0004028. 

\bibitem{quasi1} C.Csaki, J. Erlich, T. J. Hollowood, hep-th/0003020,
hep-th/0003076.

\bibitem{grs1} R. Gregory, V. A. Rubakov, S. M. Sibiryakov, hep-th/0003045. 


\bibitem{alex1} I. Olasagasti and A. Vilenkin, hep-th/0003300.

\end{references}
\end{document}